\documentclass[floats,floatfix,showpacs,amssymb,prd,twocolumn,superscriptaddress,nofootinbib,nolongbibliography,reprint]{revtex4-2}

\usepackage{amssymb,amsmath,verbatim,mathtools,needspace,enumitem,etoolbox,graphicx,physics,microtype,afterpage}

\usepackage[dvipsnames, usenames]{xcolor}

\definecolor{linkcolor}{rgb}{0.0,0.3,0.5}
\usepackage[unicode, colorlinks=true, linkcolor=linkcolor, citecolor=linkcolor, filecolor=linkcolor,urlcolor=linkcolor, pdfusetitle]{hyperref}
\usepackage[all]{hypcap}
\usepackage[T1]{fontenc}
\usepackage[utf8]{inputenc}
\usepackage{tabularx}
\allowdisplaybreaks
\renewcommand{\arraystretch}{1.4}
\newcommand{\ssim}{\mathchar"5218\relax\,}

\graphicspath{{./figure/}}

\def\apj{\rm{ApJ}}
\def\apjl{\rm{ApJ}}
\def\apjs{\rm{ApJS}}
\def\aap{\rm{A\&A}}

\def\mnras{\rm{MNRAS}}
\def\prd{\rm{PRD}}
\def\prl{\rm{PRL}}
\def\prx{\rm{PRX}}
\def\pasa{\rm{PASA}}
\def\pasp{\rm{PASP}}

\def\lrr{\rm{LRR}}

\newcommand{\jhu}{\affiliation{Department of Physics and Astronomy, Johns Hopkins University, 3400 N. Charles Street, Baltimore, Maryland 21218, USA}}
\newcommand{\bham}{\affiliation{School of Physics and Astronomy and Institute for Gravitational Wave Astronomy, University of Birmingham, Birmingham, B15 2TT, United Kingdom}}

\begin{document}

\title{Machine-learning interpolation of population-synthesis simulations\\ to interpret gravitational-wave observations: A case study}

\author{Kaze W.~K. Wong}\email{kazewong@jhu.edu} \jhu

\author{Davide Gerosa} \email{d.gerosa@bham.ac.uk} \bham

\begin{abstract}

We report on advances to interpret current and future gravitational-wave events in light of astrophysical simulations. A machine-learning emulator is trained on numerical population-synthesis predictions and inserted into a Bayesian hierarchical framework. In this case study, a modest but state-of-the-art suite of simulations of isolated binary stars is interpolated across two event parameters 
and one population parameter.
The validation process of our pipelines highlights how omitting some of the event parameters might cause errors in estimating selection effects, which propagates as systematics to the final population inference.
Using LIGO/Virgo data from O1 and O2 we infer that black holes in binaries are most likely to receive natal kicks with one-dimensional velocity dispersion $\sigma=105^{+44}_{-29}\ \rm{km/s}$. Our results showcase potential applications of machine-learning tools in conjunction with population-synthesis simulations and gravitational-wave data. \end{abstract}

\maketitle

\section{Introduction}

The observed catalog of gravitational-wave (GW) detections is growing at 
a fast pace. %
Eleven \cite{2018arXiv181112907T} (or possibly more \cite{2019arXiv190407214V}) events have been announced so far and hundreds more are expected to be observed within a few years. 
This opens a unique possibility of inferring properties of the population of merging compact binaries in the Universe.

Current state-of-the-art analyses assume that the underlying population of GW sources is described by some phenomenological parametric expression (e.g. \cite{2018arXiv181112940T}). For instance, one can assume that the mass distribution of merging binary black holes (BHs) follows a power law, and use GW data to infer its spectral index. 

As the size of the GW catalog grows, increasingly complex parametrizations and nonparametric tests (e.g. \cite{2017MNRAS.465.3254M}) will allow us to capture finer and finer details of the observed population. 

Astrophysical predictions of GW populations are typically computed using population-synthesis codes
\citep{2002MNRAS.329..897H,2017NatCo...814906S,2018MNRAS.480.2011G,2008ApJS..174..223B,2013ApJS..204...15P,2013MNRAS.431.2184G}
 --collections of prescriptions that encode our understanding and ignorance of how compact binaries form and evolve from their stellar progenitors.
The parameters needed to initialize a population-synthesis simulation are directly related to poorly understood astrophysical mechanisms ruling the lives of massive stars.
These include, for instance, efficiency of the common-envelope phase (if any), strength of the supernova kicks, fallback material, mass-loss rates, stellar winds, etc.

A possible strategy to exploit future large GW catalogs is to bypass phenomenological models and compare data directly against population-synthesis simulations. A first step in this direction consists of estimating mixing fractions between two or more precomputed models \cite{2017ApJ...846...82Z, 2017MNRAS.471.2801S,2017PhRvD..95l4046G,2019arXiv190511054B}. More ambitiously,
 one could use GW data to infer the set of code input flags that best matches the observations. This approach faces an immediate difficulty, namely that a new, computationally expensive population-synthesis simulation is required at each evaluation of the population likelihood.

Progress to overcome this limitation was recently presented by Taylor and Gerosa \cite{2018PhRvD..98h3017T}. %
By combining Gaussian process regression (GPR), principal component analysis (PCA), space-filling algorithms, and a hierarchical Bayesian framework, they were able to efficiently interpolate a precomputed bank of population-synthesis simulations and use the resulting emulator to infer the posterior distributions of the population parameters.
Their method, however, was only applied to a few simple scenarios~\cite{2018PhRvD..98h3017T}.

In this paper, we present a more realistic application: we train a GPR interpolant on a small, but state-of-the-art, set of population-synthesis predictions of BH binaries formed in isolation \cite{2018PhRvD..98h4036G}. The resulting emulator slots into a hierarchical Bayesian analysis and is fed with BH binary data from LIGO/Virgo first (O1) and second (O2) observing runs.
Under these astrophysical assumptions, we measure the natal kicks that BHs receive at birth. 

This case study sheds light on some of the challenges one has to overcome to fully compare GW data and astrophysical simulations. In particular, we show that omitting a subset of the single-event parameters, either because they are not modeled in the simulations or simply because they make the inference problem computationally prohibitive, will cause a systematic bias on the final population inference.

This paper is organized as follows. 
In Sec. \ref{pipeline} we briefly review the procedure of Ref.~\cite{2018PhRvD..98h3017T} %
and present the current application.
In Sec. \ref{sec:Results}
 we show inference results using both mock datasets and real observations. 
Finally, in Sec. \ref{sec:Discussion} we discuss some astrophysical implication of our findings and highlight future developments of this approach.

\section{Methods}
\label{pipeline}

\subsection{Hierarchical Bayesian inference} \label{sec:HBA}

For each GW event, data $d$ are routinely analyzed using Bayesian inference \cite{2018arXiv181112907T}. The chosen priors encode one's physical intuition of the underlying population and play an important role when interpreting the results \cite{2017PhRvL.119y1103V}. A hierarchical analysis aims at parametrizing the choice of prior and using data to infer the resulting ``hyperparameters.''
For clarity, in the following we will denote parameters describing single events (e.g. masses, redshifts) as ``{event parameters}'', and the hyperparameters describing the entire sample as ``{population parameters}'' (e.g. the strength of supernova kicks).
 Let us assume that a set of population parameters $\lambda$ predicts a distribution of event parameters $\theta$ %
\begin{equation}
\frac{\rm d}{{\rm d} {\theta} }r(\lambda) = r(\lambda) \,p_{\rm pop}(\theta|\lambda)\,,
\label{population}
\end{equation}
where $\int p_{\rm pop}(\theta|\lambda) {\rm d}\theta=1$ and the total rate $r(\lambda)$ is typically measured in yr$^{-1}$.
This function encodes our astrophysical assumptions on the underlying populations. It can be estimated using a parameterized model, a population-synthesis simulation, or, in our case, by evaluating a machine-learning emulator. The predicted number of events is $N(\lambda)= r(\lambda) \times T_{\rm obs}$
where $T_{\rm obs}$ is the duration of the observing run(s).

We wish to analyze a GW catalog containing $N_{\textrm{obs}}$ entries. For simplicity, we assume that all the events present in the catalog are of astrophysical origin. A more complete analysis including triggers with larger false-alarm probabilities is left to future work (e.g.~\cite{2019MNRAS.484.4008G}).
Single-event posterior $p({\theta}|{d})$ are computed using some default prior $\pi(\theta)$ which is chosen by issuers of the catalog. In practice, both prior and posterior are usually provided under the form of Monte Carlo samples~\cite{2018arXiv181112907T}. 

Detector selection effect are encoded in a function $0\leq p_{\textrm{det}}(\theta)\leq 1$, indicating the likelihood that an event with parameters $\theta$ appears in the catalog. This is used to define the observable distribution %
\begin{equation}
\frac{\rm d}{{\rm d} {\theta} }r_{\rm det}(\lambda) = r(\lambda) \,p_{\rm pop}(\theta|\lambda)\, p_{\rm det}(\theta)
\label{population}
\end{equation}
and the expected number of observations $N_{\rm det}(\lambda)= r_{\rm det}(\lambda) \times T_{\rm obs}$.
 Here we follow a common approach and approximate $p_{\textrm{det}}({\theta})$ using the single-detector semi-analytic approximation of Refs.~\cite{1993PhRvD..47.2198F,1996PhRvD..53.2878F}) as implemented in Ref.~\cite{gwdet} with a signal-to-noise ratio threshold equal to 8 and the waveform model of Ref.~\cite{2014PhRvL.113o1101H}. This was shown to be in 
 good agreement with large-scale injection campaigns~\cite{2016ApJ...833L...1A,2018arXiv181112940T}. %

All these ingredients enter the population likelihood, which has the standard expression of an inhomogeneous Poisson process (c.f. Refs.~\citep{2004AIPC..735..195L,2018PhRvD..98h3017T,2019MNRAS.486.1086M,2019PASA...36...10T} for detailed derivations). In particular, the population posterior reads
\begin{equation}
p({\lambda}|{d}) \!\propto\! \pi({\lambda})  \,e^{- N_{\rm det} ({\lambda})} N({\lambda})^{N_{\rm obs}}\! \prod_{i=1}^{N_{\rm obs}} \! \int \!\frac{p_i({\theta}|{d})}{\pi_i({\theta})} p_{\rm pop}({\theta}|{\lambda}){\rm d}{\theta} \,,
\label{eq:posterior}
\end{equation}
where $\pi(\lambda)$ is some assumed population prior. If one wishes exclude rate information from the inference, a marginalization over $N(\lambda)$ with prior $\propto 1/N(\lambda)$ yields~\cite{2018ApJ...863L..41F}
\begin{equation}
p({\lambda}|{d}) \!\propto\! \pi({\lambda})  \prod_{i=1}^{N_{\rm obs}} \! \int \!\frac{p_i({\theta}|{d})}{\pi_i({\theta})} 
\frac{p_{\rm pop}({\theta}|{\lambda})}{\int p_{\rm pop}({\theta}|{\lambda}) p_{\rm det}(\theta)}
{\rm d}{\theta} \,.
\label{eq:posteriormarg}
\end{equation}

\subsection{Training simulations} \label{sec:simulation}

In this paper we consider GW sources formed in isolation via a common-envelope phase --a leading formation channel for binary BHs (e.g. \cite{2014LRR....17....3P}). We use the set of predictions presented in Ref.~\cite{2018PhRvD..98h4036G,2019PhRvD..99j3004G} (see references therein). The simulations are performed with the \textsc{startrack} \cite{2008ApJS..174..223B} and \textsc{precession} \cite{2016PhRvD..93l4066G} codes (see also \cite{spops}).
 In particular, these runs employ the same set of assumptions used in model M10 of Ref.~\cite{2016A&A...594A..97B}, except that BH natal kicks are not suppressed compared to neutron stars. Kicks are drawn from a Maxwellian distribution with one-dimensional velocity dispersion $\sigma$. We consider seven simulations with $\sigma = $ 0, 25, 50, 70, 130, 200, and 265 km/s, the latter being the value inferred from pulsar proper motions \cite{2005MNRAS.360..974H}. 

Each event in our synthetic catalogs is described by masses, spins, redshift, as well as its contribution to the total merger rate ${\rm d}r/{\rm d}\theta$ \cite{2016ApJ...819..108B}. 
In this paper, we restrict the event parameters used in our statistical inference to masses, redshifts, and rates (see below). Even though BH spins are not directly considered, assumptions on their distribution enter the waveform, hence the detection rates. Using the various spin models developed by Ref.~\cite{2018PhRvD..98h4036G}, we verified that this indirect spin effect has a negligible impact on our final results. For concreteness, in the following we use the ``time-uniform'' model \cite{2018PhRvD..98h4036G}.
All binaries are assumed to reach the LIGO/Virgo band in quasicircular orbits.

Selection effects $p_{\rm det}(\theta)$ are computed using sensitivity curves for both LIGO in its design configuration and LIGO during O1/O2. In particular, we use the ``Design Sensitivity'' and the ``Early High Sensitivity'' from Ref.~\cite{2018LRR....21....3A}, respectively (cf.~\cite{2018arXiv181112940T}).

\subsection{Gaussian processing} \label{sec:gpr}

Astrophysical simulations are used to train a Gaussian process interpolator to quickly evaluate the rates ${\rm d}r(\lambda)/{\rm d}{\theta}$. %
 In particular, our problem has a single population parameter $\lambda=\{\sigma\}$. 
Our implementation closely follows that of Ref.~\cite{2018PhRvD..98h3017T}. We first assume a common binning scheme across all simulations and convert the  distribution using PCA.\footnote{PCA naturally allows reducing the size of the computational problem by filtering out unnecessary features \cite{2018PhRvD..98h3017T}. In this case, we are only using 7 training simulations and are able to process the entire distributions without any compression.} The resulting features are then interpolated across the hyperparameter space with GPR.

The choice of event parameters used in the inference needs to be addressed with care. A trade-off is present between the size of the vector $\theta$ and the resulting GPR accuracy.
A larger number of event parameters would increase the amount of astrophysical information captured by the analysis. However, this requires larger training banks to keep the interpolation error under control.  
This means, not surprisingly, that if we want to increase the amount of information used in the hyperparameters inference, we also need to feed in more training data to ensure the same accuracy.

We found that our set of 7 simulations allows us to accurately interpolate across source-frame chirp mass and redshift, i.e. $\theta=\{M_c,z\}$. %
 For this (admittedly modest) training bank, inserting additional parameters, like mass ratio or effective spin, significantly degrades the performance of the interpolator.
In particular, we use 40 equispaced bins in $M_c\in[5,45]$ and $z\in[0,1]$.
GPR is implemented using \textsc{scikit-learn} \cite{scikit-learn} %
with a squared exponential kernel as in Ref.~\cite{2018PhRvD..98h3017T}. %
The population posterior of Eq.~\eqref{eq:posterior} is sampled using 
\textsc{emcee} \cite{2013PASP..125..306F}.

\begin{figure}
\includegraphics[width=0.94\columnwidth]{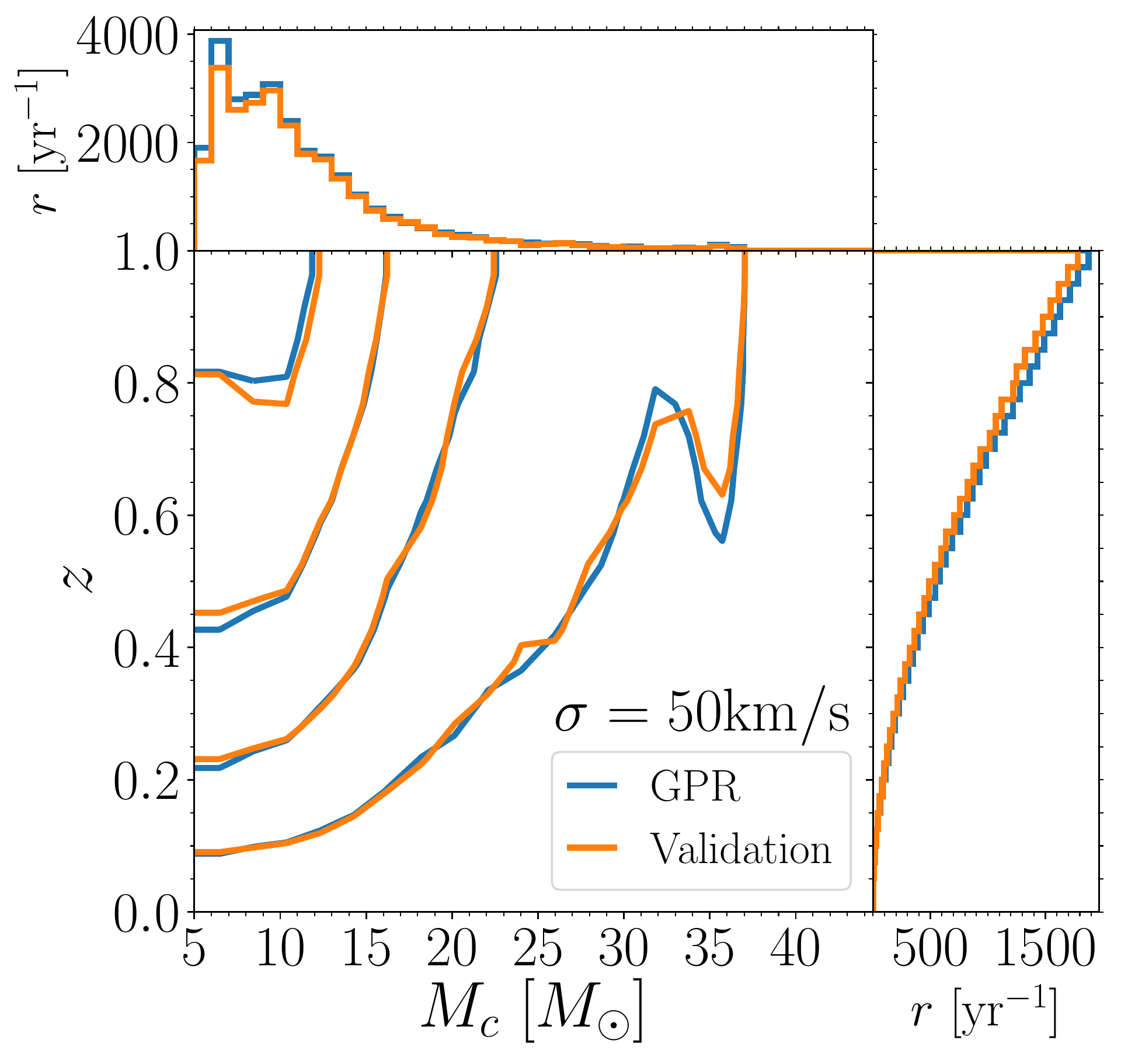}\\
\medskip
\includegraphics[width=0.94\columnwidth]{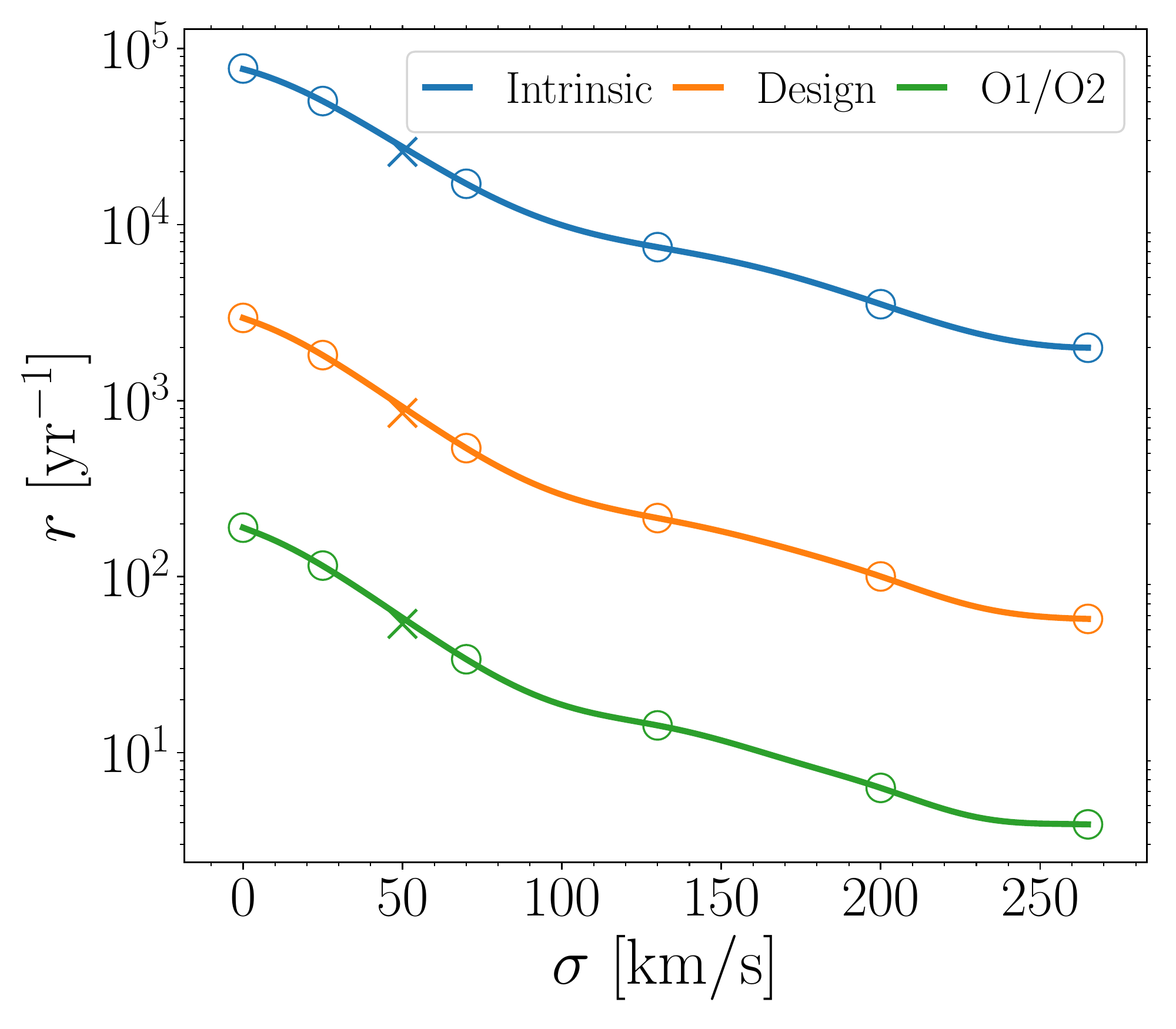}
\caption{
Out-of-sample test of our machine-learning interpolator. The simulation with $\sigma=50$ km/s is excluded from the training dataset and used to validate results. 
The top panel shows the intrinsic distributions of chirp mass and redshift ${\rm d}r/{\rm d}{\theta}$. Blue curves show the interpolated result, while orange curves show the control set. Contours mark 30\%, 50\%, 70\% and 90\% confidence intervals; side histograms show the marginalized distributions.
The bottom panel shows detection rates across the hyperparameter space. In particular, the blue line shows intrinsic rates $r$, while orange and green lines show observable rates $r_{\rm det}$ for LIGO at design sensitivity and during O1/O2, respectively. Circles mark the simulations used to train the interpolator; crosses mark the validating dataset. 
}
\label{fig:GPAccuracy_contour}
\end{figure}

We validate our pipeline using a standard out-of-sample test. 
We train our regression machine using 6 simulations and validate results against the one that was left out.
Figure \ref{fig:GPAccuracy_contour}
 shows the predicted distributions of chirp mass $M_c$, redshift $z$, and rates $r$.
Other than some small-scale differences, the interpolator accurately captures all the main features of the training set. For instance, we found a fractional difference in the intrinsic rate as small as $\Delta r/r\sim 6\%$

Reducing the number of event parameters implies that the quantities used in the inference differ from those needed to compute selection effects via $p_{\rm det}(\theta)$. Assumptions on other parameters beyond chirp mass and redshift such as mass ratio and spins will inevitably be necessary to compute waveforms. For this reason, one cannot simply interpolate ${\rm d}r(\lambda)/{\rm d}{\theta}$ across $\theta=\{M_c , z\}$ and compute $N_{\rm det}(\lambda)$ at each likelihood evaluation. We bypass the issue by computing the detectable rates from the training simulations (where mass ratios and spins are provided) and running a second GPR/PCA interpolation on ${\rm d}r_{\rm det}(\lambda)/{\rm d}{\theta}$ (cf. Fig.~\ref{fig:GPAccuracy_contour}). This approach is tested in the next section.

\section{Results} \label{sec:Results}

\begin{figure*}
\includegraphics[width=0.96\columnwidth]{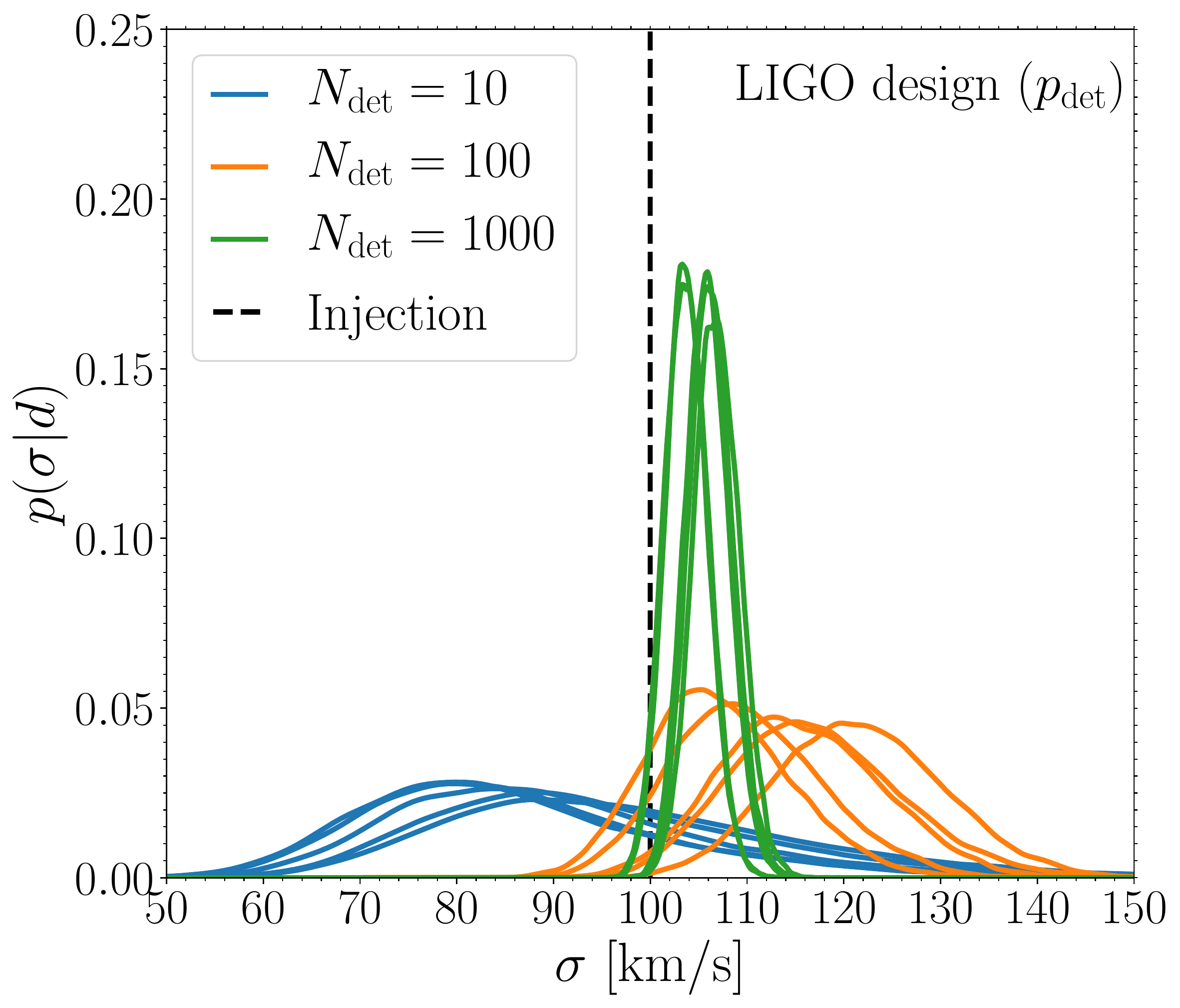}
\includegraphics[width=0.96\columnwidth]{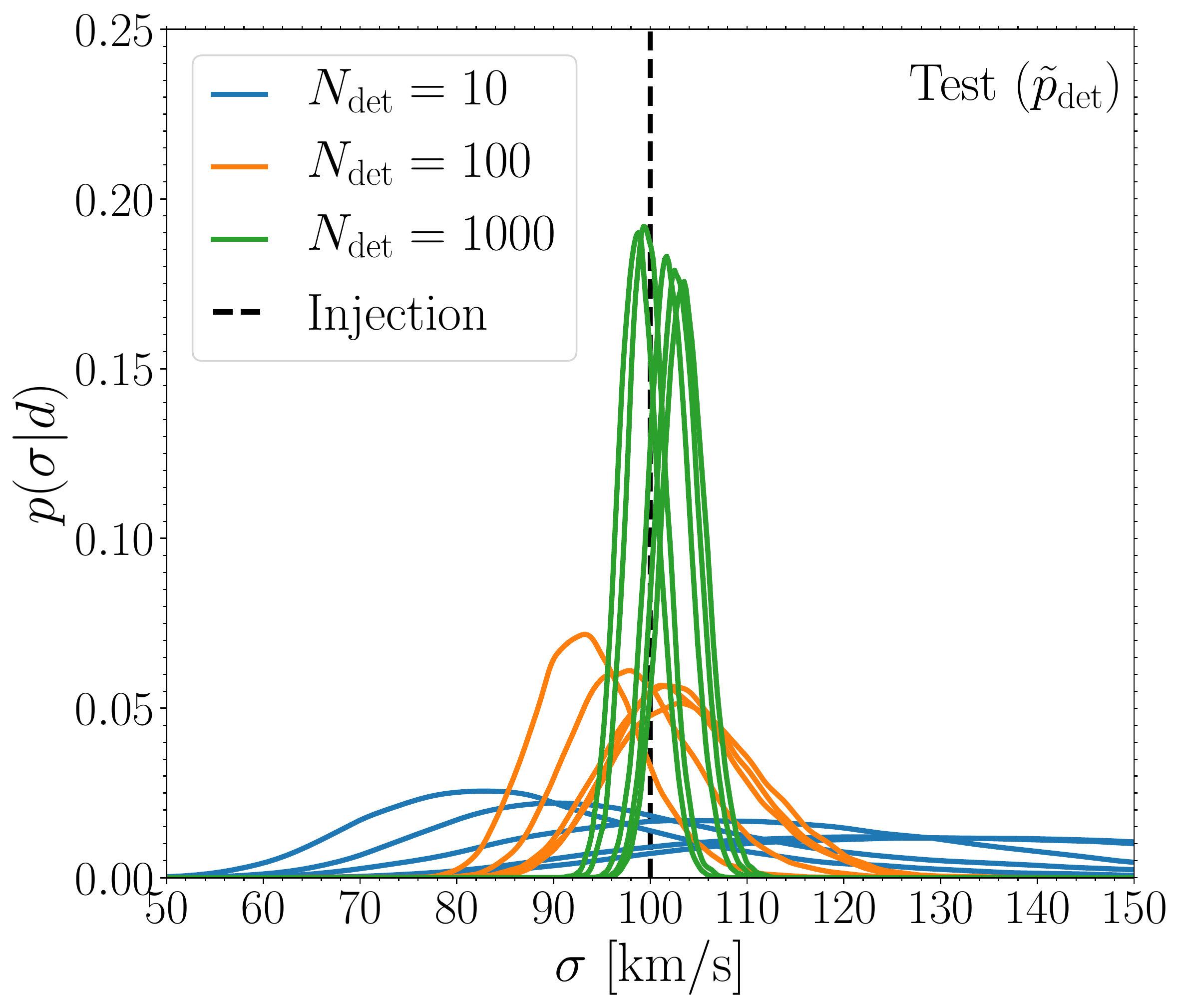}

\caption{
Injection-recovery test. The left panel considers LIGO at design sensitivity, where selection effects are included through $p_{\rm det}$. In this case, both ${\rm d}r/{\rm d}\theta$ and ${\rm d}r_{\rm det}/{\rm d}\theta$ are interpolated from population-synthesis simulations. The right panel shows a test run where we assume a fictitious LIGO detector described by $\tilde p_{\rm det}$ (see the description in the text), which allows us to only interpolate ${\rm d}r/{\rm d}\theta$. The injected value $\bar \sigma$= 100 km/s is marked by vertical dashed lines. Solid curves show posterior distributions of the strength of supernova kicks $\sigma$ assuming the predicted number of observable sources is $N_{\rm det}(\bar\sigma)=$10 (blue), 100 (orange), and 1000 (green). A few realizations are reported for each of these cases. %
} %
\label{fig:Injection}
\end{figure*}

\subsection{Mock data} \label{sec:injection}

We first apply our statistical pipeline to mock data. We assume that a population of BH binaries with true value $\bar\sigma$=100 km/s is observed by LIGO at design sensitivity.
This is a location in parameter space where we have not performed a population-synthesis simulation.
The observing time $T_{\rm obs}$ is chosen such that the predicted number of observation is $N_{\rm det}(\bar\sigma)=\{10, 100, 1000\}$.
We assume the number of entries $N_{\rm obs}$ in our mock catalogs is Poisson-distributed with mean $N_{\rm det}(\bar\sigma)$.
Injections events $\theta=\{M_c,z\}$ are generated from the ${\rm d} r_{\rm det}/{\rm d}\theta$ emulator. 
For simplicity, we assume posterior $p_i(\theta|d)$ are bivariate Gaussians centered on those extracted values with standard deviations equal to $10\%$.  
We sample the posterior of Eq.~\eqref{eq:posterior} assuming a flat priors on $M_c$, $z$ and $\sigma$.

Our injection-recovery results are shown in left panel of Fig.~\ref{fig:Injection} for several catalog realizations. %
The injected value $\bar\sigma=100$ km/s is well within the predicted posterior only for the case with $N_{\rm det}=10$. As the number of detections grows, a systematic bias becomes more and more evident. For $N_{\rm det}=1000$, the posteriors peak at $\ssim 105$ km/s and the true values lies well outside the $90\%$ confidence interval. 

This bias is somewhat expected because we are neglecting some of the event parameters and only considering $\theta=\{M_c,z\}$.
This forces us to interpolate ${\rm d}r(\lambda)/{\rm d}{\theta}$ and ${\rm d}r_{\rm det}(\lambda)/{\rm d}{\theta}$ separately.
In normal circumstance these two distributions are related by a single detectability function, i.e. ${\rm d}r_{\rm det}(\lambda)/{\rm d}{\theta} = p_{\rm det}(\theta){\rm d}r(\lambda)/{\rm d}{\theta} $.
Our pipeline, however, violates this condition because the two interpolants have different interpolation errors occurring in each bin.

We test this interpretation by considering a fictitious detector where selection effects $\tilde p_{\rm det}$ depends only on $M_c$ and $z$. This is constructed by assuming the same LIGO sensitivity curve which, however, responds to all binary BHs as if it they were equal mass and nonspinning, i.e. $\tilde p_{\rm det}(M_c, z) = p_{\rm det}(M_c, z, q=1, \mathbf{\boldsymbol\chi_1}=0, \mathbf{\boldsymbol\chi_2}=0)$. In this case, we can interpolate only ${\rm d}r/{\rm d}\theta$ using population synthesis data. Injections are constructed extracting couples $\{M_c,z\} $ from ${\rm d}r/{\rm d}\theta$ and accepting/rejecting each draw according to $\tilde p_{\rm det} (\theta)$. The factor $e^{-N_{\rm det}(\lambda)}$ in Eq.~(\ref{eq:posterior}) is estimated by integrating $ \tilde p_{\rm det}(\theta) \times {\rm d}r/{\rm d}\theta$ at each likelihood evaluation. Results are shown in the right panel 
of Fig.~\ref{fig:Injection}. We recover a largely 
unbiased estimates of the population parameter.

As discussed above, the number of parameters we can confidently interpolate is limited by size of the training dataset. %
A larger set of simulations would allow us to model more event parameters, %
consequently reducing systematic uncertainties on the resulting inference. Figure ~\ref{fig:Injection} shows, however, that the present simulation set is appropriate for $\mathcal{O}(10)$ events, as in this case statistical uncertainties largely dominates over systematics. We thus proceed by analyzing the 10 BH binary events detected during O1 and O2.

\subsection{Events from LIGO/Virgo O1 and O2}

After removing data segments contaminated by significant noise sources, LIGO/Virgo
O1 and O2 resulted in $T_{\rm obs}=48.6$ and 118 days of coincident data, respectively~\cite{2018arXiv181112907T, 2016PhRvX...6d1015A}.
We make use of posterior and prior samples of 10 binary BH coalescences publicly released by the LIGO and Virgo collaborations~\cite{2015JPhCS.610a2021V}. %
 In particular, they provide luminosity distance and detector-frame masses, which we convert to redshift and source-frame masses. A Gaussian kernel-density estimation is then employed to obtain $\pi(\theta)$ at the locations of the posterior samples, which allows approximating the integrals in Eqs.~(\ref{eq:posterior}-\ref{eq:posteriormarg}) as Monte Carlo sums.

\begin{figure}
\includegraphics[width=0.94\columnwidth]{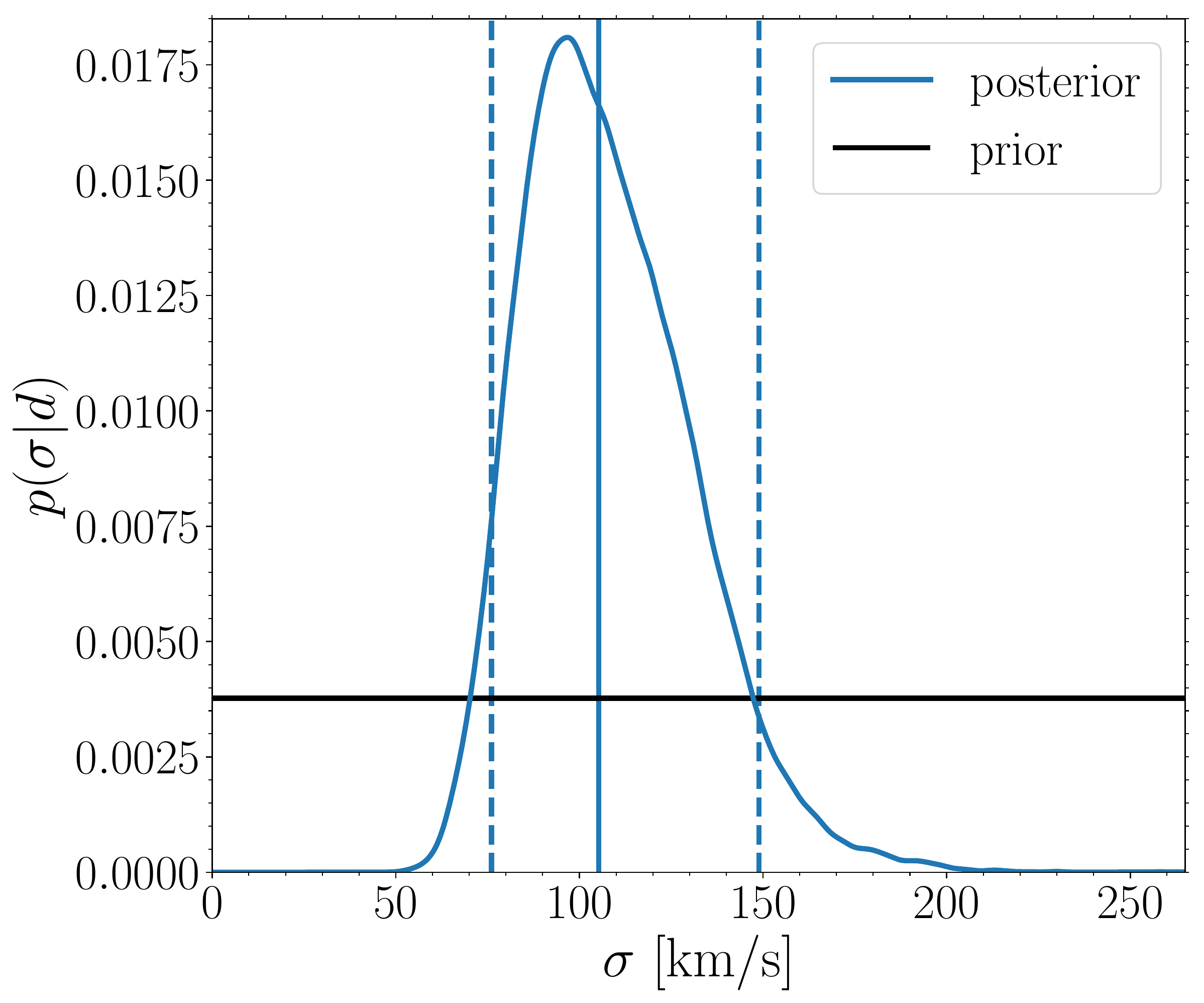}
\caption{
Constraints on the strength of BH natal kicks $\sigma$ using 10 BH binary data from LIGO/Virgo O1 and O2. The blue curve shows our posterior distribution. %
Vertical blue lines show the corresponding median (solid) and $90\%$ confidence interval (dashed). %
We assume a flat prior (horizontal black line).
}
\label{fig:MoneyPlot}
\end{figure}

The resulting inference is illustrated in Fig.~\ref{fig:MoneyPlot}, where we show the posterior distribution of the population parameter $\sigma$. This is our GW measurement of the kicks imparted to BHs at formation. Quoting median and 90\% confidence interval, we find $\sigma=105^{+44}_{-29}$ km/s. The information gain between prior and posterior, as quantified by the Kullback-Leibler divergence~\cite{kullback1951}, is $D_{\rm KL} = 1.54$.

The posterior is skewed toward high values of $\sigma$. This is because the event rate $r$ changes more (less) rapidly at low (high) values of $\sigma$ (cf. Fig. \ref{fig:GPAccuracy_contour}). Consequently, data can more (less) easily accommodate kicks that are larger (smaller) than the inferred posterior maximum.

Our inference is largely driven by the integrated event rate $r(\lambda)$. For this set of simulations, the current GW catalog does not contain enough discriminating power to perform an informative analysis restricted to $p_{\rm pop}$. Repeating our study using the marginalized likelihood of Eq.~(\ref{eq:posteriormarg}) returns a much lower information gain $D_{\rm KL} = 0.24$.

\section{Discussion} \label{sec:Discussion}

By applying the statistical framework of Ref.~\cite{2018PhRvD..98h3017T}, we analyzed current GW data using population-synthesis simulations of binary BHs formed in isolation. Simulations enter the training process of a machine-learning algorithm, in this case GPR. The resulting emulator is then used to evaluate the likelihood in the context of a standard hierarchical Bayesian analysis. The present case study puts this idea into practice using a modest set of 7 training simulations,  allowing us to showcase both prospects and pitfalls of this approach.
Overall, the pipeline returns posterior distributions of the input flags one needs to initialize and run population-synthesis simulations. These are related to poorly understood mechanisms in the lives of massive stars, like BH natal kicks, which are here measured directly from GW data.

We first tested our approach on mock catalogs and show the presence of a systematic bias that exceeds statistical uncertainties when the number of observations is $\gtrsim 100$. The omission of some event parameters causes errors when modeling detector selection effects which in turn, propagate to the population inference.
The test reported in Sec.~\ref{sec:injection} suggests that this issue can be alleviated with a larger set of training simulations, which will allow using a larger set of event parameters while keeping the interpolation error under control.
The case where some parameters need to be omitted reflects a common situation. In this study, information on mass ratio and spins are available but limited number of simulations prevented us from carrying out a more complete analysis. A more damaging scenario occurs when some variables affecting the GW signal are not modeled at all in the astrophysical simulations. Our results show the importance of developing astrophysical models where all the observables (BH spins, eccentricity, etc) are taken into account.

We also applied our procedure to BH binary data from LIGO/Virgo O1 and O2. Our results suggest that binary BHs were imparted moderate kicks at formation ($ \sigma \gtrsim 70\,{\rm km/s}$). This is in tentative agreement with position and proper-motion measurements in X-ray binaries \cite{2017MNRAS.467..298R,2016MNRAS.456..578M,2019MNRAS.489.3116A}, as well as GW measurements of BH-binary spin misalignment \cite{2017PhRvL.119a1101O,2018PhRvD..97d3014W}. Our findings are, however, in contrast with current supernova models which predict that BH natal kicks should be highly suppressed due to fallback material in the late stage of the explosion (e.g.~\cite{2012ApJ...749...91F}). The simulations used in this paper do not have a dedicated flag to tune the amount of fallback, which is instead controlled directly by the value of $\sigma$. %

We stress that our result is model dependent. This is intentional: \emph{we are interpreting GW events in light of a specific set of astrophysical assumptions}. Consequently, only those assumptions are put to test. Among the set of predictions explored here, current GW data prefer models where moderately large natal kicks are imparted onto BHs. It is natural to expect that a more complete set of training simulations might change this result qualitatively. Even assuming the same population-synthesis setup, a larger set of training simulations will allow (i) capturing degeneracies between different population parameters and (ii) efficiently interpolating across additional event parameters.%

Prior assumptions are inevitably part of any statistical analysis. Instead of relying on parametrized distributions, our approach makes use of state-of-the-art simulations in a data-driven fashion. We believe our approach presents promising avenues to infer astrophysical formation and evolutionary processes of GW sources, thus making a step forward toward the goal of GW astronomy.

\acknowledgments
We thank E.~Berti, S.~Taylor, C.~Moore, A.~Vecchio and E.~Roebber for discussions.
K.W.K.W. is supported by NSF Grant No. PHY-1841464, NSF Grant No. PHY-1912550, NSF Grant No. AST-1841358, NSF-XSEDE Grant No. PHY-090003, NASA ATP Grant No. 17-ATP17-0225, and EU H2020 Marie Skłodowska-Curie grant No. 690904. The authors would like to acknowledge networking support by the COST Action CA16104 ``GWverse''. Computational work was performed on the University of Birmingham's BlueBEAR cluster and at the Maryland Advanced Research Computing Center (MARCC). We made use of data provided by the Gravitational Wave Open Science Center, a service of LIGO Laboratory, the LIGO Scientific Collaboration, and the Virgo Collaboration.

\bibliography{sigmaspops}

\end{document}